\documentclass[aps,twocolumn,amsmath,amssymb,superscriptaddress]{revtex4-2}
\usepackage{graphicx} 
\usepackage{physics} 
\usepackage{hyperref} 

\usepackage{color}

\newcommand{\new}[1]{\textcolor{black}{#1}}

\newcommand*\subtxt[1]{_{\textnormal{#1}}}
\DeclareRobustCommand\_{\ifmmode\expandafter\subtxt\else\textunderscore\fi} 

\begin{document}

\title{Numerical analysis of a superradiance-sideband-assisted laser with a zero frequency pulling and a narrow linewidth}
\author{Mingyu Jeon}
\affiliation{Department of Physics and Astronomy \& Institute of Applied Physics, Seoul National University, Seoul 08826, Korea}
\author{Jinuk Kim}
\affiliation{Korea Research Institute of Standards and Science, Daejeon 34113, Korea}
\author{Kyungwon An}
\affiliation{Department of Physics and Astronomy \& Institute of Applied Physics, Seoul National University, Seoul 08826, Korea}
\date{\today}

\begin{abstract}
Numerical simulations based on the quantum Langevin equations have been performed for a large number of two-level atoms in a beam interacting with a low-Q cavity with the atomic initial superposition states close to the north pole of the Bloch sphere.
When the pump Rabi frequency was modulated at $\Delta\_{pa}$ with zero pump-atom detuning for various cavity-atom detunings, we obtained 
a lasing peak at the atomic resonance and superradiant lasing peaks at $\pm\Delta\_{pa}$ simultaneously while the central peak exhibiting a zero frequency pulling coefficient. 
The linewidth of the central peak was reduced beyond the gain narrowing as the mean number of atoms was increased, resulting in a minimum linewidth as small as a millionth of the atomic or cavity linewdith.
A pump carrier detuning caused asymmetric heights for the side superradiant peaks, the height difference of which can be used to lock the pump laser to the atom within the linewidth of the central lasing peak. 
Our results may lead to development of a new type of ultra-stable active optical clocks for future frequency standards when applied to proper atomic systems.
\end{abstract}

\maketitle

\section{Introduction}

The precise measurement of time has been one of most important endeavors in the development of science.
Many modern technologies including communications, computer networks and GPS rely on precise measurements of time in order to operate properly\cite{review-article}.
Accurate and stable clocks also help advancing fundamental sciences. A test of Lorentz symmetry has been performed using two ytterbium optical clocks with precision in the order of $10^{-18}$\cite{Sanner2019}. More precise time measurement would lead to new discoveries as well as new technologies and enable many fundamental sciences which might be impossible otherwise\cite{Aeppli2024}.

A great deal of efforts are made to create better clocks\cite{PhysRevLett.123.033201}, such as clocks using quantum logic technology\cite{doi:10.1126/science.1114375,doi:10.1126/science.1154622}, single-ion\cite{Micke2020}, rare-earth atoms cooled and localized in an optical lattice\cite{PhysRevLett.89.230801,PhysRevLett.91.173005,Takamoto2005}, etc.
An idea of an active optical clock based on superradiance has also been pursued\cite{Bohnet2012} and its initial tests showed a potential to surpass the passive optical clocks\cite{doi:10.1126/sciadv.1601231} such as the optical lattice clocks employing a narrow-linewidth laser locked to a ultra-stable reference cavity and stabilized to an ultra-narrower atomic transition. 
In an active optical clock, the atomic system itself acts as an oscillator and generates radiation of a narrow linewidth without relying on an external cavity\cite{Zhang2023}. 

In a typical superradiant laser operating as an active optical clock\cite{Bohnet2012}, the atoms with an ultra-narrow linewidth are trapped in a bad cavity and excited by a pulsed pump laser in a superradiant state repeatedly. Because the atomic linewidth $\gamma\_a$ is much narrower than the cavity linewdith $\gamma\_c$, the so-called frequency pulling of the resulting superradince lasing of frequency $\omega\_{sr}$ is given by $\delta \omega\equiv \omega\_{sr}-\omega\_c \simeq \frac{\gamma\_a}{\gamma\_c}(\omega\_c-\omega\_a)$, thereby achieving a negligible frequency-pulling coefficient in the order of $\frac{\gamma\_a}{\gamma\_c}\ll 1$, where $\omega\_{a(c)}$ is the atomic(cavity) resonance frequency. This is an advantage of a superradiant active optical clock over the passive optical clocks that need be stabilized to an external cavity, which is prone to thermal noise, and thus limiting the clock's ultimate precision\cite{PhysRevLett.116.013001,Beloy2021,Schioppo2017,Grotti2018,PhysRevA.98.053443}. 

The superradiant laser at its present form has a shortcoming though, that is, the averaging time of the frequency is limited because of its pulsed operation.
In order to overcome this shortcoming, a quasi-superradiant laser employing a dense beam of two-level atoms traversing a bad cavity has been proposed for a continuous-wave operation\cite{PhysRevLett.125.253602}. The bad cavity condition enforces the total atomic polarization to be proportional to the field amplitude, analogously to the rate equation approximation, and as a result, the system exhibites superradiance-like behavoir in the steady state although the atoms are fully inverted initially without any atomic coherence, 
The resulting cavity field shows both the linewidth and the frequency pulling coefficient decreasing as the mean number of atoms, and consequently the number of collective emission events, in the cavity is increased. 

The remaining frequency pulling coefficient in the work of Ref.~\cite{PhysRevLett.125.253602}, although very small, would eventually limit the performance of the active optical clock based on this principle. By noting that the atomic superposition state can coherently inject coherence to the field and thus that the frequency of the superradiance is independent of the cavity-atom detuning\cite{Kim2022}, one may wonder what happens if the atoms are initially prepared with both the population inversion and the atomic coherence by using a superposition state close to the north pole of the Bloch sphere. 

The present study is motivated by this question. Through numerical simulations, we have found that a proper initial superposition state results in simultaneous operation of an ordinary lasing at the atomic resonance and superradiant lasing at symmetrically detuned frequencies at $\pm\Delta\_{pa}$ when the amplitude of the pump laser for preparing the initial state was sinusoidally modulated at $\Delta\_{pa}$. Moreover, the ordinary lasing exhibited both linewidth narrowing and a zero frequency pulling coefficient because of the symmetric interaction of the superradiant lasing components with the ordinary lasing one. The zero frequency pulling took place over the cavity-atom detuning comparable to the cavity linewidth, which is assumed to be much larger than that of the atoms in the bad cavity limit.  For a non-zero pump carrier detuning, the superradiant lasing components showed asymmetric peak heights, the difference between which was proportional to the pump carrier detuning. By using this asymmmetry, a feedback loop can be implemented to nullify the remaining pump carrier detuning. We found the expected signal-to-noise ratio of the lasing signal would be enough to eliminate the pump frequency fluctuations through the feedback, and as a result, the ultimate linewidth of the central ordinary lasing peak can be as narrow as $10^{-6}$ of the natural linewidth of the atoms. Our simulation parameters such as a 50-MHz cavity decay rate and a 0.5-MHz atom-cavity coupling constant are comparable with high-density high velocity rare-earth ion beams such as Ca$^+$ with about 400 nm transition wavelengths. Our results can be tailored for a specific atom system and can be used as a guideline for building an actual active clock system. 

This paper is organized as follows. In Chapter II, we briefly recapitulate the quantum Langevin equations formulated by H. Liu et al. \cite{PhysRevLett.125.253602} with our own comments on some technical issues related to the derivation.  We then introduce three different configurations in the superposition-state pumping that are considered in our simulations and provide the expressions for the corresponding initial states for the atoms. We also explain the tilted atomic beam geometry and its physical consequence as well as how the pump laser linewidth is included and how the power spectral density is calculated. Simulation results are discussed in Chapter III, where we first reproduce the results of Ref.~\cite{PhysRevLett.125.253602} in order to verify the validity of our code. We then present the results with a non-modulated pump laser with a nonzero detuning exciting an initial superposition state close to the north pole of the Bloch sphere for each atom. The results motivate us to perform simulations with a modulated pump laser with a zero carrier detuning. The results with an ordinary lasing component occuring at the atomic resonance with two side bands of superradiant lasing are presented. The linewidth narrowing of the ordinary lasing component as a function of the mean number of atoms in the cavity is discussed. In Chapter IV, we consider the effect of a non-zero pump carrrier detuning and discuss a feedback scheme to nullify the overall shift of the three-peak spectrum using the asymmetric heights of the superradiant peaks. In Chapter V, we summarize our work and remark on possible extension of the present work for actual experimental studies. 

\section{Theory}
 
\subsection{Equations of Motion}
We consider a laser made of a beam of two-level atoms and a cavity. The atoms are initially excited in a common superposition state of the ground and excited states before they enter the cavity. We assume an atomic beam crossing the cavity mode at a small tilt angle so as to ensure traveling-wave atom-cavity interaction\cite{Choi2006,Hong2012} with the atom-cavity coupling constant uniform inside the cavity. 
We employ the same set of differential equations reported by H. Liu {\em et al.}\cite{PhysRevLett.125.253602}. Their formalism is recapitulated below. 

Our system is described by the Tavis-Cummings Hamiltonian\cite{Tavis-Cummings} in the rotating frame of the atom as, 
\begin{equation}
\hat{H}/\hbar={\Delta}{\hat{a}}^{\dagger}{\hat{a}}+g{\sum\limits_j}\eta({\textbf{x}_j})({\hat{\sigma_j}}^{+}{\hat{a}}+{\hat{a}}^{\dag}{\hat{\sigma_j}^{-}}),
\label{eq1}
\end{equation}
where $\Delta=\omega\_c-\omega\_a$ is the cavity-atom detuning, $\hat{a}(\hat{a}^\dag)$ is the photon annihilation(creation) operator, $\hat{\sigma}_j^-(\hat{\sigma}_j^+)$ is the lowering(raising) operator for the $j$th atom, $g$ is the atom-cavity coupling constant (half width), also known as the vacuum Rabi frequency, and $\eta({\textbf{x}_j})$ is the normalized cavity mode function evaluated at the position $\textbf{x}_j$ of the $j$th atom ($|\eta({\textbf{x}_j})|\le 1$).
Without including the loss of the cavity, time derivatives of the field and atomic spin operators can be obtained as
\begin{eqnarray}
\frac{d\hat{a}}{dt}&=&\frac{i}{\hbar}[\hat{H},\hat{a}] =-i\Delta\hat{a}-ig\hat{J}^{-},\label{eq2}\\
\frac{d{\hat{a}^{\dagger}}}{dt}&=&\frac{i}{\hbar}[\hat{H},\hat{a}^\dag] =i{\Delta}{\hat{a}^{\dagger}}+ig{\hat{J}^{+}}, \label{eq3}\\
\frac{d{\hat{\sigma}^{-}_{j}}}{dt}&=&\frac{i}{\hbar}[\hat{H},\hat{\sigma}_j^-] =ig{\eta_{j}}{\hat{\sigma}^{z}_{j}}{\hat{a}}, \label{eq4}\\
\frac{d{\hat{\sigma}^{+}_{j}}}{dt}&=&\frac{i}{\hbar}[\hat{H},\hat{\sigma}_j^+] =-ig{\eta_{j}}{\hat{a}^{\dagger}}{\hat{\sigma}^{z}_{j}}, \label{eq5}\\
\frac{d{\hat{\sigma}^{z}_{j}}}{dt}&=&\frac{i}{\hbar}[\hat{H},\hat{\sigma}_j^z] =2ig{\eta_{j}}({\hat{a}^{\dagger}}{\hat{\sigma}^{-}_{j}}-{\hat{\sigma}^{+}_{j}}{\hat{a}}),\label{eq6}
\end{eqnarray}
where $\hat{J}^{\pm}\equiv \sum_{k=1}^N \eta({\textbf{x}_k})\hat{\sigma}_k^\pm$ and the relations $[\hat{a},\hat{\sigma}_j^\pm]=0=[\hat{a}^\dag,\hat{\sigma}_j^\pm]$ and $\hat{\sigma}^{z}_{j}=[\hat{\sigma}^{+}_{j},\hat{\sigma}^{-}_{j}]$ are used. 
When we introduce the cavity decay with a decay rate $\kappa$ (full width), Eqs.~(\ref{eq2}) and (\ref{eq3}) are modified as
\begin{eqnarray}
\frac{d\hat{a}}{dt}&=&-(\kappa/2+i\Delta)\hat{a} -ig \hat{J}^- -\sqrt{\kappa}\hat{\xi}, \label{eq7}\\
\frac{d\hat{a}^\dag}{dt}&=&-(\kappa/2-i\Delta)\hat{a}^\dag+ig\hat{J}^+ -\sqrt{\kappa}\hat{\xi}^\dag, \label{eq8}
\end{eqnarray}
where we also include the noise terms with $\hat{\xi}(\hat{\xi}^\dag)$ the noise operator associated with $\hat{a}(\hat{a}^\dag)$. The noise terms arise from the coupling of $\hat{a}$ and $\hat{a}^\dag$ to the vacuum outside through the cavity mirrors. 
We assume a bad cavity, for which $\kappa\gg \sqrt{\bar{N}}g$ is satisfied with $\bar{N}$ the mean number of atoms in the cavity. Under this condition, Eqs.~(\ref{eq7}) and (\ref{eq8}) quickly reach a steady state, so we can approximately set $\frac{d\hat{a}}{dt}=0=\frac{d\hat{a}^\dag}{dt}$. The bad cavity approximation, similar to the rate equation approximation in quantum optics, is expected to give a correct steady-state solution, which we are interested in. It allows us to express $\hat{a}$ and $\hat{a}^\dag$ as
\begin{eqnarray}
\hat{a}&=&-\frac{g}{\Delta-i\kappa/2}\hat{J}^{-}+\frac{i\sqrt{\kappa}}{\Delta-i\kappa/2}\hat{\xi} \nonumber\\
&=&-\frac{1}{2g}(\Gamma_\Delta+i\Gamma\_c)\left(\hat{J}^{-}-\frac{2i}{\sqrt{\Gamma_0}}\hat{\xi}\right) , \label{eq9}\\
\hat{a}^{\dagger}&=&-\frac{g}{\Delta+i\kappa/2}\hat{J}^{+}-\frac{i\sqrt{\kappa}}{\Delta+i\kappa/2}\hat{\xi}^{\dagger} \nonumber\\
&=&-\frac{1}{2g}(\Gamma_\Delta-i\Gamma\_c)\left(\hat{J}^{+}+\frac{2i}{\sqrt{\Gamma_0}}\hat{\xi}^\dag\right), \label{eq10}
\end{eqnarray}
where 
\begin{equation}
\Gamma_\Delta \equiv \frac{2g^2\Delta}{\Delta^2+(\kappa/2)^2}, ~ \Gamma\_c \equiv \frac{g^2\kappa}{\Delta^2+(\kappa/2)^2}, ~
\Gamma_0 =\frac{4g^2}{\kappa}.
\label{eq10'}
\end{equation}

Substituting Eqs.~(\ref{eq9}) and (\ref{eq10}) into Eqs.~(\ref{eq4}) - (\ref{eq6}), we then obtain
\begin{eqnarray}
\frac{d \hat{\sigma}^-_j}{dt}&=&-\frac{i}{2}\eta_j (\Gamma_\Delta+i\Gamma\_c)\hat{\sigma}^z_j 
\left(\hat{J}^- -\frac{2i}{\sqrt{\Gamma_0}}\hat{\xi}\right), \label{eq11}\\
\frac{d \hat{\sigma}^+_j}{dt}&=&\frac{i}{2}\eta_j (\Gamma_\Delta-i\Gamma\_c)
\left(\hat{J}^+ +\frac{2i}{\sqrt{\Gamma_0}}\hat{\xi}^\dag\right)\hat{\sigma}^z_j,\label{eq12}\\
\frac{d \hat{\sigma}^z_j}{dt}&=& -i \eta_j \left[ (\Gamma_\Delta -i \Gamma\_c)
\left(\hat{J}^+ +\frac{2i}{\sqrt{\Gamma_0}} \hat{\xi}^\dag \right)\hat{\sigma}_j^- \right. \nonumber\\
& &~~~~\left.-(\Gamma_\Delta +i \Gamma\_c)\hat{\sigma}_j^+
\left(\hat{J}^- -\frac{2i}{\sqrt{\Gamma_0}} \hat{\xi} \right) 
\right],\label{eq13}
\end{eqnarray}
from which we can get the differential equations for $\hat{\sigma}_j^x=\hat{\sigma}_j^+ + \hat{\sigma}_j^-$, $\hat{\sigma}_j^y=-i(\hat{\sigma}_j^+ - \hat{\sigma}_j^-)$ and $\hat{\sigma}_j^z$ as
\begin{eqnarray}
\frac{d \hat{\sigma}_j^x}{dt}&=&\frac{\Gamma\_c}{2}\eta_j \left[\hat{J}^x \hat{\sigma}_j^z -\eta_j \hat{\sigma}_j^x (\hat{\sigma}_j^z+1) \right]\nonumber\\
& &-\frac{\Gamma_\Delta}{2}\eta_j \left[\hat{J}^y \hat{\sigma}_j^z -\eta_j \hat{\sigma}_j^y (\hat{\sigma}_j^z+1)\right] \nonumber\\
& & -\frac{\Gamma\_c}{\sqrt{\Gamma_0}} \eta_j \hat{\sigma}_j^z \hat{\xi}^p 
-\frac{\Gamma_\Delta}{\sqrt{\Gamma_0}} \eta_j \hat{\sigma}_j^z \hat{\xi}^q, \label{eq14}\\
\frac{d \hat{\sigma}_j^y}{dt}&=&\frac{\Gamma\_c}{2}\eta_j \left[ \hat{J}^y \hat{\sigma}_j^z -\eta_j \hat{\sigma}_j^y (\hat{\sigma}_j^z+1) \right]\nonumber\\
& &+\frac{\Gamma_\Delta}{2}\eta_j \left[\hat{J}^x \hat{\sigma}_j^z -\eta_j \hat{\sigma}_j^x (\hat{\sigma}_j^z+1)\right] \nonumber\\
& &+\frac{\Gamma\_c}{\sqrt{\Gamma_0}} \eta_j \hat{\sigma}_j^z \hat{\xi}^q 
-\frac{\Gamma_\Delta}{\sqrt{\Gamma_0}} \eta_j \hat{\sigma}_j^z \hat{\xi}^p,
\label{eq15}\\
\frac{d \hat{\sigma}^z_j}{dt}&=& -\frac{\Gamma\_c}{2} \eta_j (\hat{J}^x \hat{\sigma}_j^x+\hat{J}^y \hat{\sigma}_j^y +2\eta_j \hat{\sigma}_j^z)\nonumber\\
& &+\frac{ \Gamma_\Delta }{2}\eta_j (\hat{J}^y \hat{\sigma}_j^x-\hat{J}^x \hat{\sigma}_j^y +2i \eta_j \hat{\sigma}_j^z )\nonumber\\
& &+\frac{\Gamma\_c}{\sqrt{\Gamma_0}} \eta_j (\hat{\xi}^p \hat{\sigma}_j^x -\hat{\xi}^q  \hat{\sigma}_j^y)\nonumber\\
& &+\frac{\Gamma_\Delta}{\sqrt{\Gamma_0}} \eta_j (\hat{\xi}^q \hat{\sigma}_j^x +\hat{\xi}^p \hat{\sigma}_j^y),
\label{eq16}
\end{eqnarray}
where $\hat{\xi}^q=\hat{\xi}^\dag +\hat{\xi}$ and $\hat{\xi}^p=-i(\hat{\xi}^\dag -\hat{\xi})$.
The stochastic differential equations used in the numerical simulation are then obtained by applying the semiclassical approximations $\hat{J}^{x(y)}\rightarrow J^{x(y)}$ and $\hat{\sigma}_j^{x(y)}\rightarrow s_j^{x(y)}$, where $J^{x(y)}$ and $s_j^{x(y)}$ are c-numbers.
\begin{eqnarray}
\frac{d s^x_j}{dt} &=&\frac{\Gamma\_c}{2}  \eta_j [J^x s^z_j -\underline{\eta_j s^x_j (s^z_j+1)}]\nonumber\\
& &- \frac{\Gamma_\Delta}{2}\eta_j [J^y s^z_j -\underline{\eta_j s^y_j(s^z_j+1)}] \nonumber\\
& &~~~    -\frac{\Gamma\_c}{\sqrt{\Gamma_0}}\eta_j s^z_j \xi^p+\frac{\Gamma_\Delta}{\sqrt{\Gamma_0}}\eta_j s^z_j \xi^q,
\label{eq17}\\
\frac{d s^y_j}{dt} &=& \frac{\Gamma\_c}{2}\eta_j[J^y s^z_j -\underline{\eta_j s^y_j(s^z_j+1)}]\nonumber\\
& &+\frac{\Gamma_\Delta}{2}\eta_j[J^x s^z_j-\underline{\eta_j s^x_j (s^z_j+1)}] \nonumber \\
& &~~~ +\frac{\Gamma\_c}{\sqrt{\Gamma_0}}\eta_j s^z_j \xi^q-\frac{\Gamma_\Delta}{\sqrt{\Gamma_0}}\eta_j s^z_j \xi^p,
\label{eq18}\\
\frac{d s^z_j}{dt} &=& -\frac{\Gamma\_c}{2}\eta_j (J^x s^x_j+J^y s^y_j+\underline{2\eta_j  s^z_j}) \nonumber\\
  & &  +\frac{\Gamma_\Delta}{2}\eta_j(J^y s^x_j-J^x s^y_j+\underline{2i\eta_j s_j^z}) \nonumber\\
& &~~~   + \frac{\Gamma\_c}{\sqrt{\Gamma_0}}  \eta_j (s^x_j \xi^p-s^y_j \xi^q) \nonumber\\
 & &~~~~  +\frac{\Gamma\_c}{\sqrt{\Gamma_0}}\eta_j(s^x_j \xi^q+s^y_j \xi^p). 
\label{eq19}
\end{eqnarray}
The terms underlined above come from the commutation relation between $\hat{J}^q$ and $\hat{\sigma}_j^p$ ($q, p=\pm,x,y,z$). If we apply the semiclassical approximations directly to Eqs.~(\ref{eq11})-(\ref{eq13}) and use $J^\pm=(J^x\pm iJ^y)$ to obtain the differential equations for $s_j^{x,y,z}$, the underlined terms would be absent because the c-numbers commute with each other. Since $|J^q|\gg |s_j^p|$ for $N\gg 1$, the underlined terms can be neglected unless we deal with a small number of atoms in the cavity.

\subsection{Superposition-State Pumping}

H. Liu {\em et al.} \cite{PhysRevLett.125.253602} studied the case where the atoms are initially prepared in the upper level of lasing without atomic coherence before they enter the cavity. The atomic beam was assumed to cross the cavity mode perpendicularly, so the atom-cavity coupling constant was varied as $\eta_j g$ with $\eta_j$ the mode function for $j$th atom.  In the present study, we are interested in what happens if atoms are initially prepared in a common superposition state for all atoms with a uniform atom-cavity coupling constant. We consider three different pumping cases for the initial superposition states. 

\subsubsection{Superposition state with a nonzero pump-atom detuning}\label{sec.II-B-1}

The pump laser for preparing the superposition state is detuned from the atomic resonance by $\Delta\_{pa}\ne 0$. The pumping time is assumed to be much shorter than the atomic damping time, so we can neglect the atomic damping during the pumping process. The initial atomic state can be calculated with the optical Bloch equation given by
\begin{equation}
\begin{split}
\Dot{s_x}&=\Delta\_{pa} s_y,\\
\Dot{s_y}&=-\Delta\_{pa} s_x+\Omega s_z,\\
\Dot{s_z}&=-\Omega s_y,
\end{split}
\label{eq20}
\end{equation}
in the rotating frame of the pump field for $t_0\le t \le t_0 + \tau\_p$. 
The atom is assumed to enter the pump field at $t=t_0$ and leave the pump field at $t=t_0+\tau\_p$. 
With the initial condition $s_x(t_0)=0, s_y(t_0)=0, s_z(t_0)=-1$ and $\Omega \gg \Delta_{\mathrm{pa}}$ with a top-hat distribution for the pump Rabi frequency $\Omega$ in time, we obtain
\begin{equation}
\begin{split}
s_x^0&\simeq -\sin(\Omega \tau\_p)\sin(\Delta\_{pa} \tau\_p),\\
s_y^0&\simeq -\sin(\Omega \tau\_p)\cos(\Delta\_{pa} \tau\_p),\\
s_z^0&\simeq -\cos(\Omega \tau\_p).
\end{split}
\label{eq21}
\end{equation}
Once the atom leaves the pump field, it enters the cavity and interacts with the cavity field there, generating new frequency components in the spectrum of the field according to Eqs.~(\ref{eq17})-(\ref{eq19}), which are expressed in the rotating frame of the atom. Therefore, we need to 
transform the above result back to the rotating frame of the atom at $t=t_0+\tau\_p$:
\begin{equation}
\begin{split}
s_x^0&\rightarrow s_x^0 \cos\Delta\_{pa}(t_0+\tau\_p)-s_y^0\sin\Delta\_{pa}(t_0+\tau\_p) \\
&~~~~~~~= \sin(\Omega \tau\_p)\sin(\Delta\_{pa} t_0),\\
s_y^0&\rightarrow s_x^0\sin\Delta\_{pa}(t_0+\tau\_p)+s_y^0 \cos\Delta\_{pa}(t_0+\tau\_p) \\
&~~~~~~~= - \sin(\Omega \tau\_p)\cos(\Delta\_{pa} t_0),\\
s_z^0&\rightarrow s_z^0=-\cos(\Omega \tau\_p).
\end{split}
\label{eq21'}
\end{equation}
Note that the pump phase at time $t_0$ is encoded in $s_x^0$ and $s_y^0$. 
Using the relation $s_z^0=\rho\_{ee}-\rho\_{gg}=2\rho\_{ee}-1$ with $\rho\_{ee}(\rho\_{gg})$ the initial excited(ground)-state population, Eq.~(\ref{eq21'}) can be rewritten as
\begin{equation}
\begin{split}
s_x^0&\simeq 2 \sqrt{ \rho\_{ee}(1-\rho\_{ee})}\sin(\Delta\_{pa} t_0),\\
s_y^0&\simeq -2 \sqrt{ \rho\_{ee}(1-\rho\_{ee})}\cos(\Delta\_{pa} t_0),\\
s_z^0&\simeq 2 \rho\_{ee}-1.
\end{split}
\label{eq22}
\end{equation}
The results of H. Liu {\em et al.} \cite{PhysRevLett.125.253602} can be reproduced with $\rho\_{ee}=1, \Delta\_{pa}=0$ and random $\eta_j$'s as shown in Fig.~\ref{fig1}.

\subsubsection{The case with two opposite pump-atom detunings}\label{sec-II-B-2}

In this case, the atoms are pumped with two opposite detunings, $\pm \Delta\_{pa}$. The pump field can be expressed as $\frac{1}{2}E_0 \left\{ \exp[ -i(\omega\_a+\Delta\_{pa})t] +\exp[-i(\omega\_a-\Delta\_{pa})t]\right\}=E_0 \cos(\Delta\_{pa}t)\exp(-i\omega\_a t)$.
Under the rotating wave approximation in the frame of the atom, it is as if the pump is resonant with the atoms with its Rabi frequency modulated as $\Omega \cos(\Delta\_{pa}t)$ with $\Omega \gg \Delta\_{pa}$.
So, the Bloch equation can be written as
\begin{equation}
\begin{split}
\dot{s_x}&=0,\\
\dot{s_y}&= \Omega \cos(\Delta\_{pa}t) s_z,\\
\dot{s_z}&=-\Omega \cos(\Delta\_{pa}t) s_y.
\end{split}
\label{eq23}
\end{equation}
Differentiating $\dot{s_z}$ equation and substituting $s_y$ and $\dot{s_y}$ expressions there, we obtain
\begin{equation}
\Ddot{s_z}+\Delta\_{pa}\tan(\Delta\_{pa}t)\dot{s_z}+\Omega^2\cos^2(\Delta\_{pa}t)s_z=0.
\label{eq24}
\end{equation}
We have solved Eq.~(\ref{eq24}) with the initial condition $s_z(t_0)=-1, s_x(t_0)=s_y(t_0)=0$,
for which the atom is assumed to enter the pump field at $t=t_0$ and leave the pump field at $t=t_0+\tau\_p$. 
The atomic state after the pump is obtained as
\begin{equation}
\begin{split}
s_x^0&=0.\\
s_y^0&=-\sin\left\{\frac{\Omega}{\Delta\_{pa}}\left[\sin \Delta\_{pa}(t_0+ \tau\_p)-\sin(\Delta\_{pa} t_0)\right]\right\},\\
s_z^0&=-\cos\left\{\frac{\Omega}{\Delta\_{pa}}\left[\sin \Delta\_{pa}(t_0+\tau\_p)-\sin(\Delta\_{pa} t_0)\right]\right\}.
\end{split}
\label{eq25}
\end{equation}
With $\Delta\_{pa}\tau\_p\ll 1$, Eq.~(\ref{eq25}) can be approximated as
\begin{equation}
\begin{split}
s_x^0&=0.\\
s_y^0&\simeq-\sin\left[ \Omega\cos(\Delta\_{pa} t_0) \tau\_p \right],\\
s_z^0&\simeq-\cos\left[ \Omega\cos(\Delta\_{pa} t_0) \tau\_p \right].
\label{eq26}
\end{split}
\end{equation}
Note that the pump phase at time $t_0$ is encoded in $s_y^0$ and $s_z^0$. 

\subsubsection{The case with two opposite pump-atom detunings with a common offset}\label{sec-II-B-3}

Here we are interested in the effect of overall pump detunings. The pump detunings are given by $\pm{\Delta}_{\mathrm{pa}}+\delta$, respectively, where $\delta<<\Delta_{\mathrm{pa}}$ is a common offset or the average of two detunings. The Bloch equation can be written in the rotating frame of the pump carrier frequency ($\omega\_a+\delta$) as
\begin{equation}
\begin{split}
\dot{s_x}&=\delta\cdot{s_y}\\
\dot{s_y}&=-\delta\cdot{s_x}+ \Omega \cos(\Delta\_{pa}t)s_z\\
\dot{s_z}&=-\Omega \cos(\Delta\_{pa}t)s_y
\end{split}
\label{eq27}
\end{equation}
Assuming $\Omega\gg \delta$, we can apply the same approximation as the one used in Eq.~(\ref{eq21}) to obtain
\begin{equation}
\begin{split}
s_x^0&\simeq-\sin\left\{\frac{\Omega}{\Delta\_{pa}}\left[\sin \Delta\_{pa} (t_0+\tau\_p)-\sin(\Delta\_{pa} t_0)\right]\right\}\sin(\delta\tau\_p),\\
s_y^0&\simeq-\sin\left\{\frac{\Omega}{\Delta\_{pa}}\left[\sin \Delta\_{pa} (t_0+\tau\_p)-\sin(\Delta\_{pa} t_0)\right]\right\}\cos(\delta\tau\_p),\\
s_z^0&\simeq-\cos\left\{\frac{\Omega}{\Delta\_{pa}}\left[\sin \Delta\_{pa}(t_0+\tau\_p)-\sin(\Delta\_{pa} t_0)\right]\right\}.
\end{split}
\label{eq28}
\end{equation}
For $\Delta\_{pa}\tau\_p\ll 1$, Eq.~(\ref{eq28}) is further simplified as
\begin{equation}
\begin{split}
s_x^0&\simeq-\sin\left[\Omega\cos(\Delta\_{pa}t_0) \tau\_p \right]\sin(\delta \tau\_p),\\
s_y^0&\simeq-\sin\left[\Omega \cos(\Delta\_{pa}t_0) \tau\_p\right]\cos(\delta \tau\_p),\\
s_z^0&\simeq-\cos\left[\Omega\cos(\Delta\_{pa}t_0)\tau\_p \right].
\end{split}
\label{eq31}
\end{equation}
We should transform this result back to the rotating frame of the atom at $t=t_0+\tau\_p$ as we did in Eq.~(\ref{eq21'}). Comparing Eq.~(\ref{eq21}) and Eq.~(\ref{eq31}), we find the correspondence $\Omega\leftrightarrow \Omega\cos(\Delta\_{pa}t_0), \Delta\_{pa}\leftrightarrow \delta$. Therefore, the result in the rotating frame of the atom is 
\begin{equation}
\begin{split}
s_x^0&\simeq \sin\left[\Omega\cos(\Delta\_{pa}t_0) \tau\_p \right]\sin(\delta t_0),\\
s_y^0&\simeq-\sin\left[\Omega \cos(\Delta\_{pa}t_0) \tau\_p\right]\cos(\delta t_0),\\
s_z^0&\simeq-\cos\left[\Omega\cos(\Delta\_{pa}t_0)\tau\_p \right].
\end{split}
\label{eq32}
\end{equation}
We recover Eq.~(\ref{eq26}) when $\delta=0$.

\subsubsection{Correspondence between $s_\alpha$ and $s_j^\alpha$ with $\alpha=x,y,z$}

It should be noted that the Bloch vector components $s_x, s_y$ and $s_z$ correspond to $s_j^x, s_j^y$ and $s_j^z$ in Eqs.~(\ref{eq17})-(\ref{eq19}), respectively.
The Bloch vector in Eqs.~(\ref{eq20}), (\ref{eq23}) and (\ref{eq27}) is defined in the rotating frame of atom
\begin{equation}
\begin{split}
s_{x}&=2\Re[\rho\_{eg}],\\
s_{y}&=-2\Im[\rho\_{eg}],\\
s_{z}&=\rho\_{ee}-\rho\_{gg},
\end{split}
\end{equation}
where $\rho\_{eg}$ is the off-diagonal element of the density matrix. Note
\begin{equation}
\begin{split}
\langle \hat{\sigma}^x\rangle={\rm Tr}[\hat{\sigma}^x\rho]=\rho\_{ab}+\rho\_{ba}=2\Re[\rho\_{ab}]=s_x,\\
\langle \hat{\sigma}^y\rangle={\rm Tr}[\hat{\sigma}^y\rho]=i\rho\_{ab}-i\rho\_{ba}=-2\Im[\rho\_{ab}]=s_y,\\
\langle \hat{\sigma}^z\rangle={\rm Tr}[\hat{\sigma}^z\rho]=\rho\_{aa}-\rho\_{bb}=s_z, 
\end{split}
\end{equation}
and by the semiclassical approximation, we have 
$\hat{\sigma}_j^x\rightarrow s_j^x$, $\hat{\sigma}_j^y\rightarrow s_j^y$ and $\hat{\sigma}_j^z\rightarrow s_j^z$. Therefore, we can identify our Bloch vector components $s_x, s_y$ and $s_z$ correspond to $s_j^x, s_j^y$ and $s_j^z$ in Eqs.~(\ref{eq17})-(\ref{eq19}), respectively.

\subsection{Simulation Methods} \label{secII-C} 

We incorporate the projection noise of spin operators in the initial values of the Bloch vector components in a way similar to the Monte-Carlo simulation. The values
of $s_x^0 , s_y^0$ and $s_z^0$ for each incident atom are randomly chosen to be either -1 or +1 
while their average value are given by Eqs. (\ref{eq22}), (\ref{eq26}) or (\ref{eq32}), depending on the pumping conditions. 

The linewidth $\Delta\omega\_p$ of the pump laser is included in terms of an additional random-fluctuating phase term in the phase of the pump laser. In Eqs. (\ref{eq22}), (\ref{eq26}) or (\ref{eq32}), the pump phase $\Delta\_{pa}t_0$ is replaced with $\Delta\_{pa} t_0+\phi\_{noise}$, where the value of $\phi\_{noise}$ is given by a Gaussian distribution, the standard deviation of which equals the square root of \hypertarget{Q1}{\new{$\Delta\omega\_p\tau\_p$}}. 

The residual Doppler shift of the atomic beam in the direction of the cavity mode is considered in terms of the atomic transverse velocity $v\_{tr}$, which is given by the Maxwell-Boltzmann velocity distribution centered at $v\_{tr}=\overline{v\_{tr}}$ with a width $\Delta v\_{tr}=\frac{\delta\_D}{k}$ with $k=\omega\_c/c$. During the interaction time, the position of the atom changes at the speed of $v\_{tr}$, which has no effect in our simulation assuming the traveling-wave atom-cavity interaction and thus a uniform coupling constant. But it introduces a Doppler shift in the cavity-atom detuning as $\Delta=\omega\_c-\omega\_a -kv\_{tr}$ in Eqs. (\ref{eq17}), (\ref{eq18}) and (\ref{eq19}) through $\Gamma_\Delta$.

The injection rate of atoms in the cavity is adjusted to obtain a desired mean number of atoms in the cavity. Each new atom interacts with the cavity for the atom-cavity interaction time $\tau$ and then it is removed from the calculations of Eqs. (\ref{eq17}), (\ref{eq18}) and (\ref{eq19}) and excluded from the summations for $J^x$ and $J^y$.
Integration of Eqs. (\ref{eq17}), (\ref{eq18}) and (\ref{eq19}) are done utilizing the 4th-order Runge-Kutta method.
After each time step, $J^x$ and $J^y$ are evaluated.
Due to the bad cavity assumption, $\hat{a}$ and $\hat{a}^{\dagger}$ are proportional to $\hat{J}^-$ and $\hat{J}^+$ as shown in Eqs. (\ref{eq9}) and (\ref{eq10}).
This allows us to calculate the first-order correlation $g^{(1)}(\tau)$ of the cavity field and its spectrum in terms of  $\hat{J}^-$ and $\hat{J}^+$ instead of $\hat{a}$ and $\hat{a}^{\dagger}$: $g^{(1)}(\tau)\propto \langle \hat{J}^+(t+\tau)\hat{J}^-(t) \rangle\simeq \overline{{J}^+(t+\tau) {J}^-(t)}$, where the bar indicates an average over time $t$.
The spectrum of the cavity field is then obtained by the Fourier transform of  $g^{(1)}(\tau)$.

\section{Simulation Result}

\begin{figure}
\centering\includegraphics[width=0.5\textwidth]{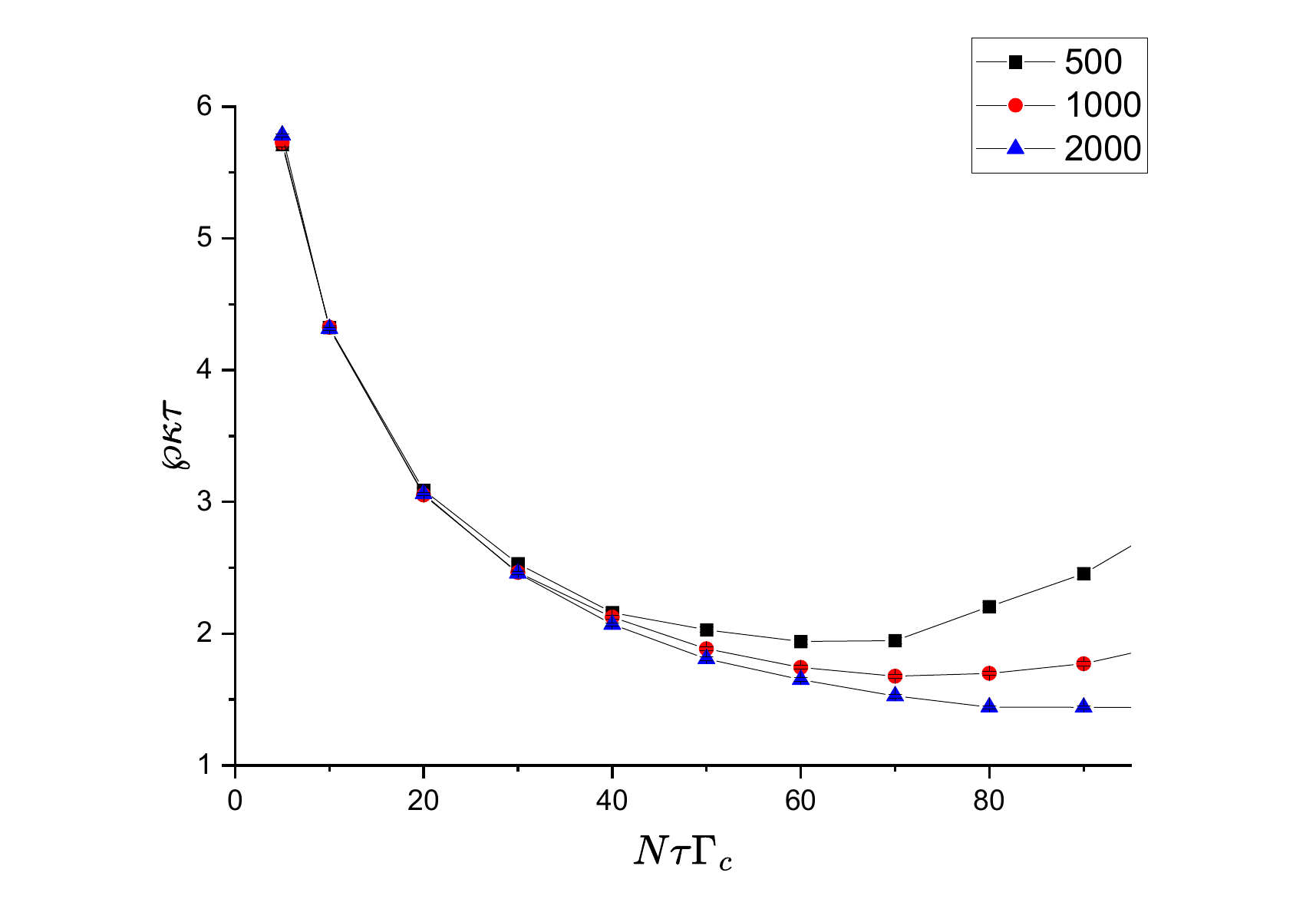}
\caption{{\bf Reproduction of the result by H. Liu {\em et al.}\cite{PhysRevLett.125.253602} for checking the validity of our code.} Simulation conditions, identical to those in Ref.~\cite{PhysRevLett.125.253602}, are as follows: the cavity decay rate (full width) is $\kappa = 2\pi\times 50$ MHz, the interaction time is $\tau = 20~{\mu}$s, the residual Doppler shift is $\delta\_D/2\pi=0.1/\tau$ and the number $N$ of atoms inside the cavity is 500, 1,000 or 2,000.
The horizontal axis is given by $N\tau\Gamma\_c$, where $\Gamma\_c=\frac{4 g^2}{\kappa}$ for zero pump-atom detuning as defined in Eq.~(\ref{eq10'}). The value of  $N\tau\Gamma\_c$, varied by $g$ in the simulation, corresponds to the number of collective emission events during the interaction time. The vertical axis $\cal{P}\kappa\tau$ corresponds to the normalized pulling coefficients.
Our result is consistent with that in Ref.~\cite{PhysRevLett.125.253602}.}
\label{fig1}
\end{figure}

Before performing our numerical studies, we tested our code by reproducing the already-published results by H. Liu {\em et al.}\cite{PhysRevLett.125.253602}. Our simulation result is shown in Fig.~\ref{fig1}, well matching that in Ref.\cite{PhysRevLett.125.253602}, and thus approving the validity of our code.

\begin{figure*}
\centering\includegraphics[width=0.59\textwidth]{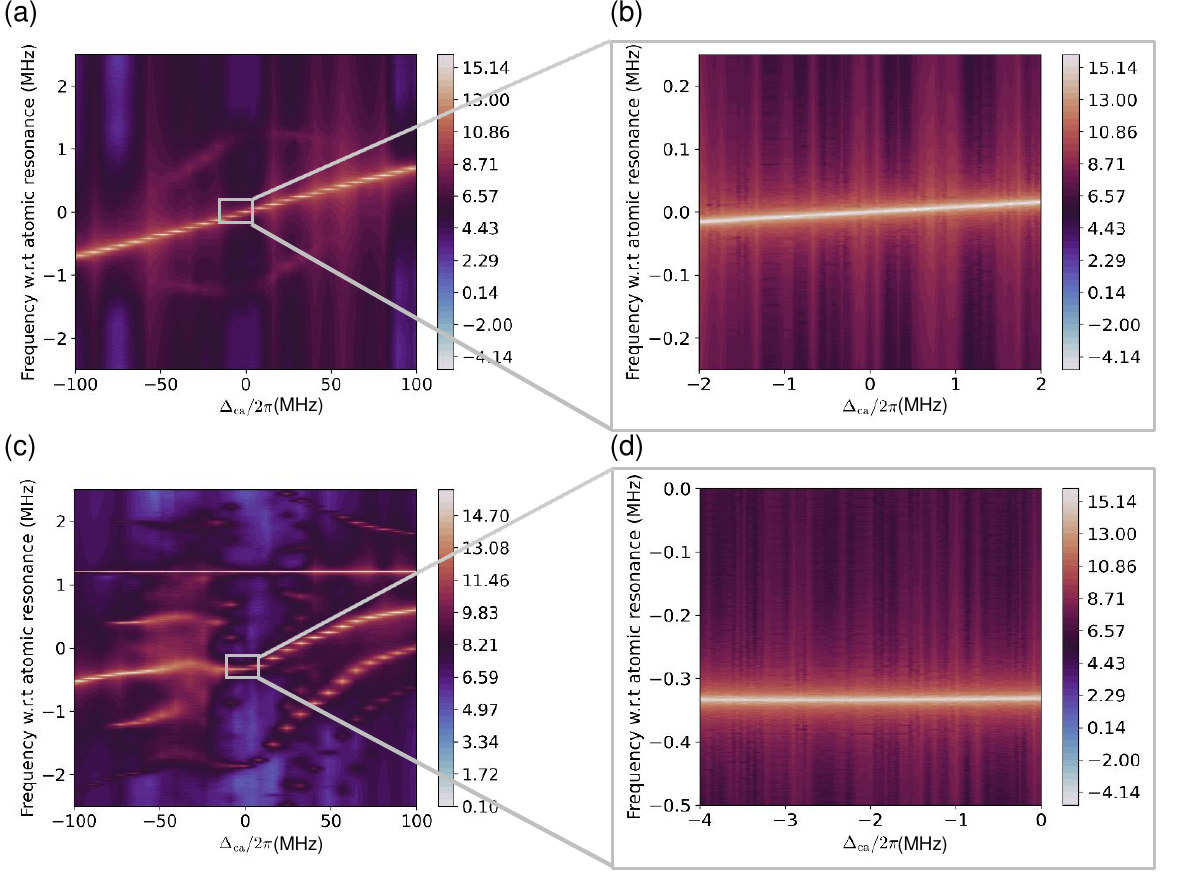}
\caption{{\bf Comparison of on-resonance population pumping and superposition-state pumping with nonzero pump-atom detuning.}  (a) Power spectral density of the intracavity field with the initial conditions, $\Omega\tau\_p=\pi$ ($\rho\_{ee}=1$) and $\Delta\_{pa}=0$ in Eq~(\ref{eq22}). (b) Magnified view of the region enclosed by a rectangle in (a), showing a weak frequency pulling effect as a function of $\Delta_{ca}$. (c) Power spectral density of the intracavity field with the initial conditions, $\Omega\tau\_p=0.7952\pi$ ($\rho\_{ee}=0.8$) and $\Delta\_{pa}/2\pi=1.4$ MHz in Eq~(\ref{eq22}). (d) Magnified view of the rectangular region in (c), exhibiting the central lasing frequency independent of the cavity-atom detuning $\Delta\_{ca}$ but with a small offset. It is due to the interaction of the ordinary lasing component with the superradiant lasing component (the horizontal line in the power spectral density plot) occurring at the pump frequency.
Common simulation parameters: $\kappa/2\pi=50$ MHz (full width), $g/2\pi=0.25$ MHz (half width), $\tau=1.0~\mu$s, $\delta\_D/2\pi=0.1/\tau$ and $N=2,000$. Each power spectral density is plotted in the log scale, color coded in the scale bar on the right.
}
\label{fig2}
\end{figure*}

\begin{figure*}
\centering\includegraphics[width=0.78\textwidth]{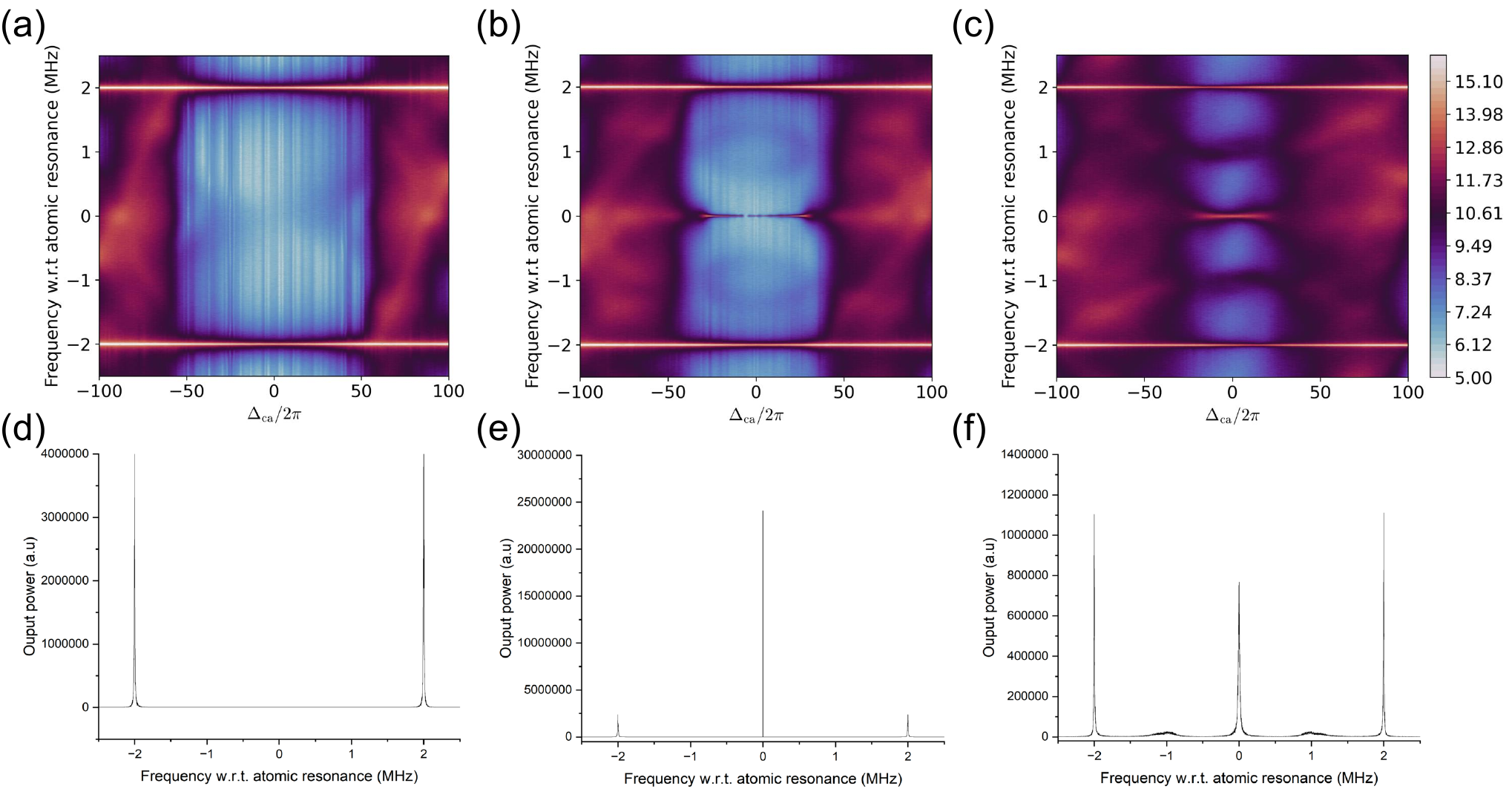}
\caption{{\bf Spectral densities under the superposition-state pumping with two opposite pump-atom detunings.} Common simulation conditions: the pump-atom detunings at $\pm 2.0$ MHz with zero pump carrier detuning $\delta=0$, $\Omega/2\pi=12$ MHz, $\tau\_p=0.0414~{\mu}$s ($\Omega\tau\_p=0.936\pi$),
$\kappa/2\pi=50$ MHz, $g/2\pi=0.25$ MHz, $\delta\_D/2\pi=0.1/\tau$ and $N=10,000$. Spectral densities of the cavity field as a function of the cavity-atom detuning $\Delta\_{ca}$ in the steady state are shown in (a) with the atom-cavity interaction time $\tau=0.36 ~\mu$s, in (b) with $\tau=0.40 ~\mu$s, and in (c) with $\tau=0.44~\mu$s, respectively. Spectra at $\Delta\_{ca}=0$ in each case are shown in (d), (e) and (f), respectively.
The horizontal line segment indicated by an ellipse in (b) is the ordinary lasing line exhibiting zero frequency pulling effect. It is fixed at the atom frequency, unaffected by the cavity-atom detuning over a significant range comparable to the cavity linewidth ($\sim$50 MHz). The horizontal lines at $\pm 0.8$ MHz in the power spectral density plots correspond to the superradiant lasing occurring at the pump frequencies. }
\label{fig3}
\end{figure*}

\begin{figure*}
\centering\includegraphics[width=0.75\textwidth]{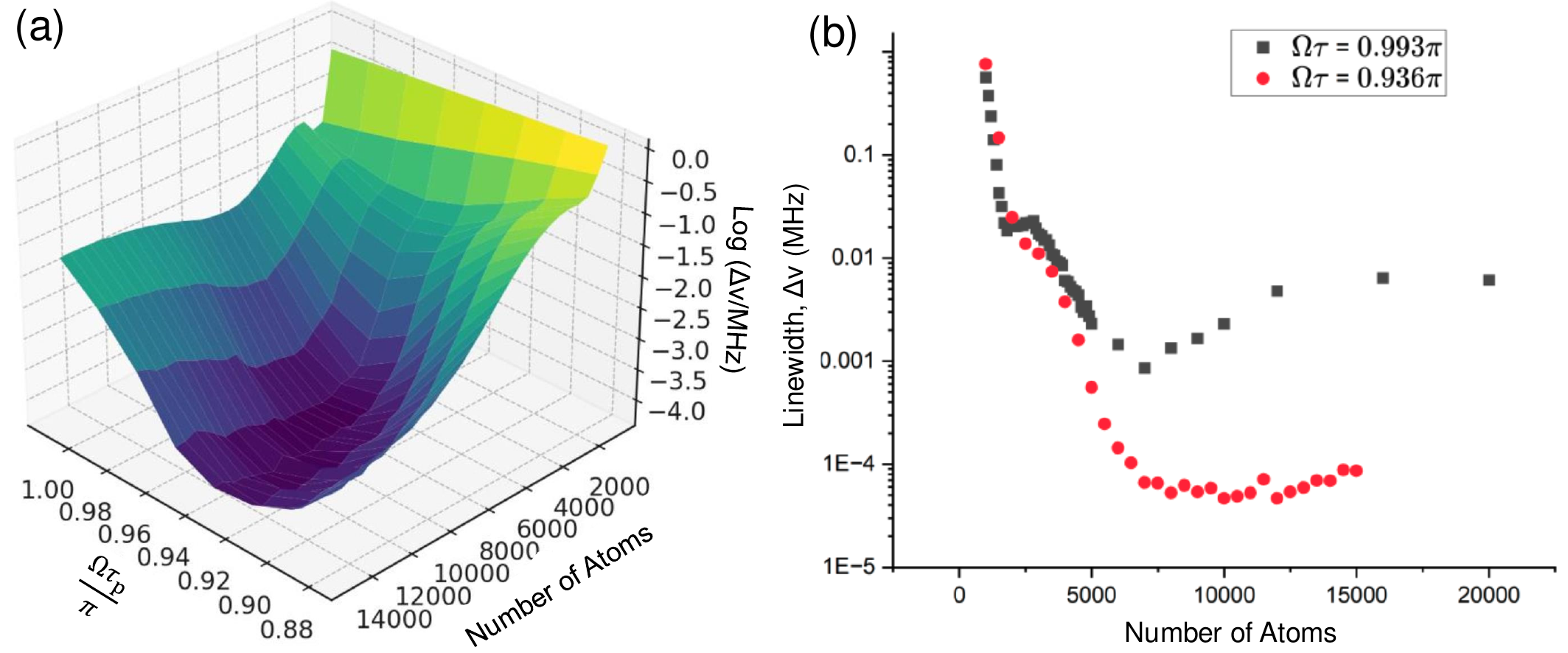}
\caption{{\bf Linewidth of the central peak in the spectrum under the superposition-state pumping with two opposite pump-atom detunings ($\delta=0$).} (a) Linewidth in MHz in log scale as a function of the pump pulse area $\Omega\tau\_p$ and the mean number $N$ of atoms in the cavity. The minimum occurs when $\Omega\tau\_p\simeq 0.94\pi$ and $N\simeq 10,000$ under our simulation conditions listed below. (b) Cross sections of the surface plot in (a) at $\Omega\tau\_p=0.936\pi$ and $0.993\pi$. For the former, the minimum linewidth is about 50 Hz, amounting to 1\% of the collective emission rate $\Gamma\_c=4g^2/\kappa$, corresponding to the number of collective emission of $N\Gamma\_c\tau\simeq 130$. Simulation conditions are $\kappa/2\pi=50$ MHz, $g/2\pi=0.25$ MHz, $\tau=0.4~{\mu}$s, $\delta\_D/2\pi=0.1/\tau$, $\Delta\_{pa}/2\pi=\pm 2$  MHz and $\Omega/2\pi=12$ MHz. The pump pulse area is varied from 0.879$\pi$ to 1.02$\pi$ by changing the pumping time $\tau\_p$ from 0.0366 $\mu$s to 0.0422 $\mu$s. 
}
\label{fig4}
\end{figure*}

\begin{figure}
\includegraphics[width=0.5\textwidth]{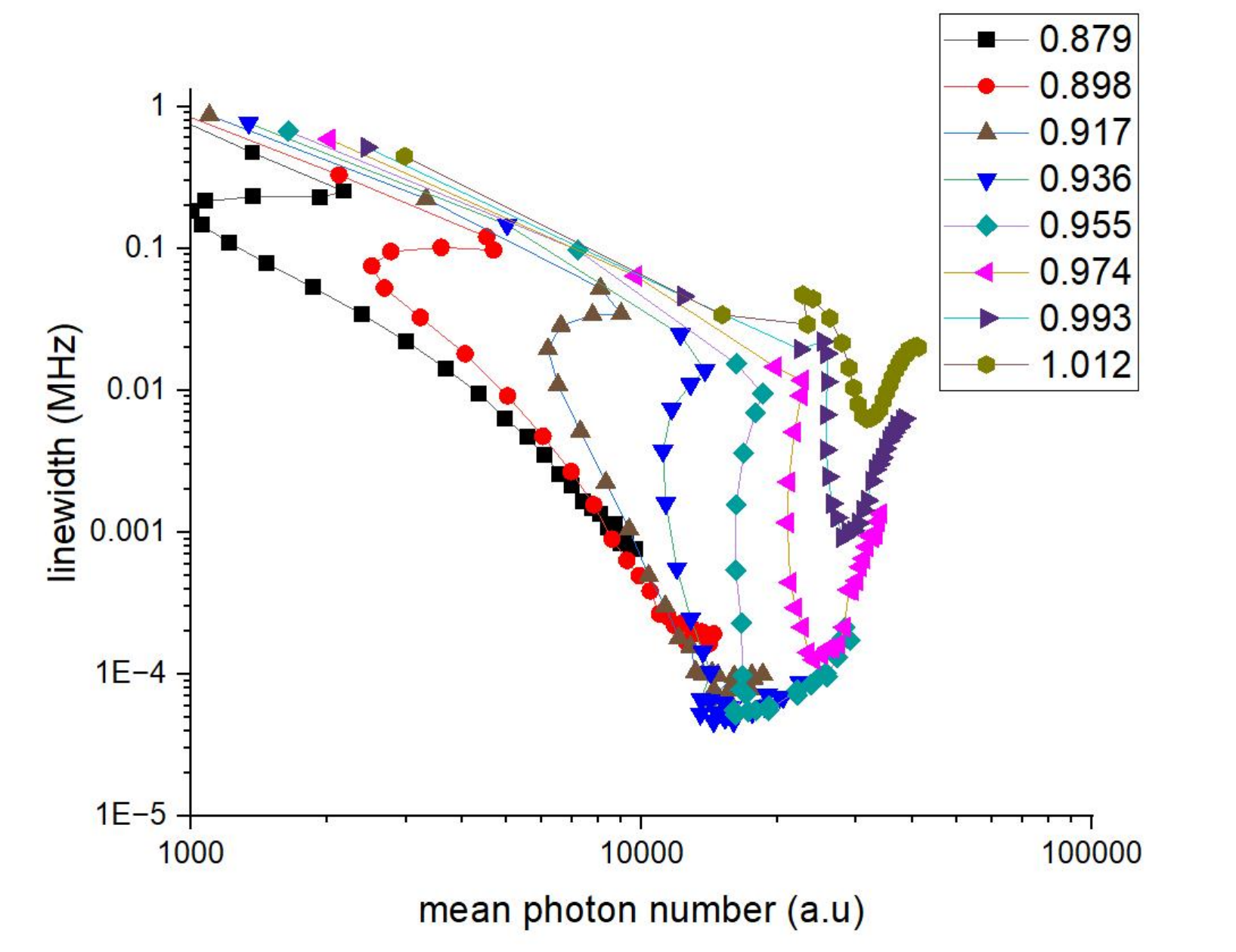}
\caption{{\bf Linewidth of the central peak vs. the intracavity mean photon number associated with the central peak for various $\Omega\tau\_p$.}
For the pump pulse area $0.92\pi<\Omega\tau\_p<\pi$, the linewidth of the central peak deceases, inversely proportional to the mean photon number, until it reaches the pump linewidth. This is nothing but the gain narrowing in the conventional lasers. Beyond the pump linewidth, as the gain increases, the mean photon number does not change but the linewidth is further decreased to reach its minimum. The linewidth starts to increase as the gain is further increased beyond the minimum as the mean photon number also increases. The smallest linewidth of 50 Hz is obtained when $\Omega\tau\_p=0.936\pi$ and $N\simeq10,000$. 
}
\label{fig5}
\end{figure}

\begin{figure*}
\centering\includegraphics[width=0.7\textwidth]{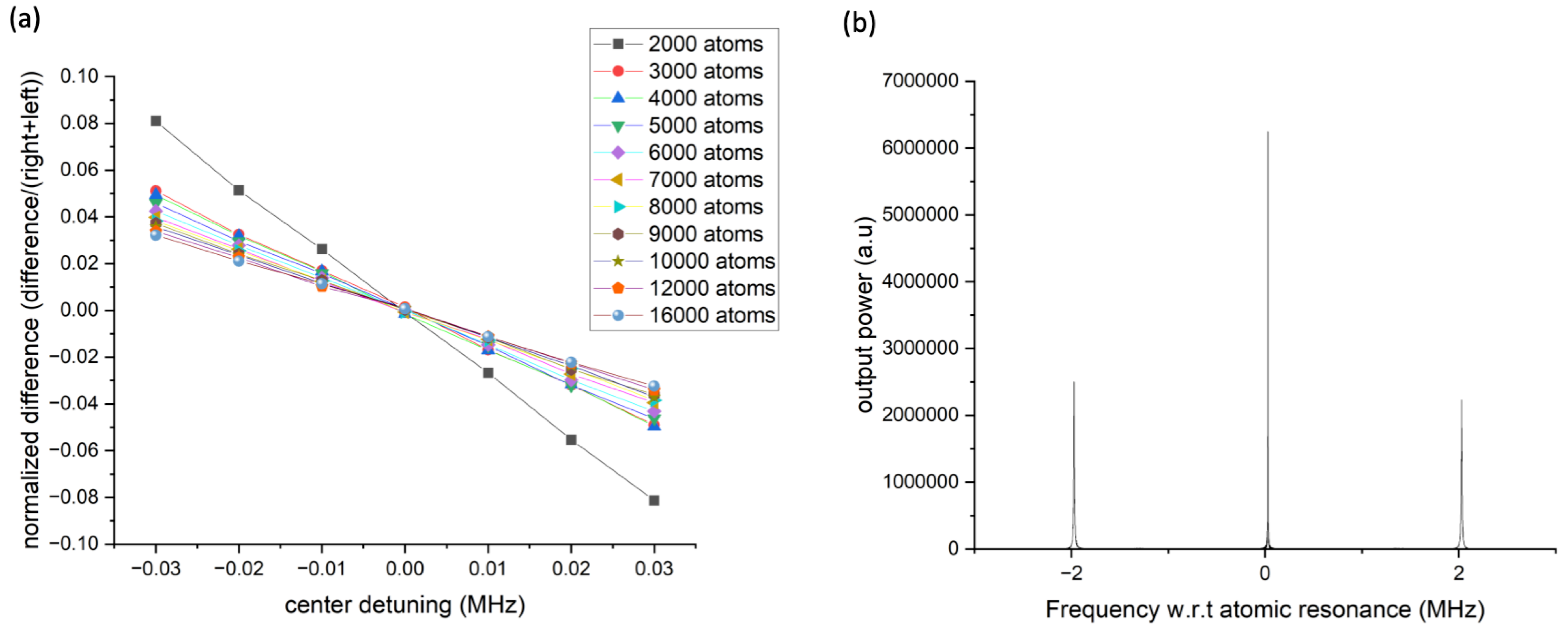}
\caption{{\bf The effect of the pump carrier detuning $\delta$ from the atomic resonance.} Nonzero pump carrier detuning introduces an overall shift of the spectrum by $\delta$ as well as an imbalance between the superradiant peak heights.
(a) Normalized difference or contrast ratio defined as (peak difference)/(peak sum) as a function of the carrier detuning for various mean number of atoms in the cavity. (b) Spectrum of the cavity field for $\delta/2\pi=0.03$ MHz and $N=10,000$.
Simulation conditions are 
$\kappa/2\pi=50$ MHz, $g/2\pi=0.25$ MHz, $\tau=0.4~{\mu}$s, $\delta\_D/2\pi=0.1/\tau$, $\Delta\_{pa}/2\pi=\delta \pm 2$ MHz, $\tau\_p=0.04138~{\mu}$s and $\Omega/2\pi=12$ MHz ($\Omega\tau\_p=0.993\pi$). 
}
\label{fig6}
\end{figure*}

We then moved to the simulations with the initial superposition state with nonzero pump-atom detuning corresponding to the first case discussed in Sec.~\ref{sec.II-B-1}. When $s_x^0$ or $s_y^0$ is nonzero in the initial superposition state, $J^x$ and $J^y$ can build up directly from initial nonzero $s_j^x$ and $s_j^y$ in Eqs.~(\ref{eq17}), (\ref{eq18}) and (\ref{eq19}). Since
$s_j^x\propto \sin(\Delta\_{pa}t_j)$ and $s_j^y\propto \cos(\Delta\_{pa}t_j)$ according to Eq.~(\ref{eq22}) with $t_j$ the arrival time of the $j$th atom at the pump field, the pump phase is encoded in the individual atomic initial states and thus $\langle \hat{a}^\dag\rangle\propto J^x+iJ^y$ oscillates as $e^{-i\Delta\_{pa}t}$ in time, resulting in a superradiant lasing at the pump frequency. If $|s_x^0|$ and $|s_y^0|$ are much smaller than $s_z^0\sim 1$, an ordinary lasing due to population inversion can also take place, and 
we can have both the ordinary lasing and the superradiant lasing as shown in Fig.~\ref{fig2}(b). If $s_z^0$ significantly deviates from unity, we only get the superradiant lasing.

When both the superradiant lasing and the ordinary lasing occur simultaneously, because of the cross terms such as $J^x s_j^x$ and $J^y s_j^y$ in $\dot{s_j^z}$ equation, Eq.~(\ref{eq19}), the frequency of the ordinary lasing is affected by that of the superradiant lasing, resulting in its frequency shift. When we compare Fig.~\ref{fig2}(a) showing the ordinary lasing spectrum and Fig.~\ref{fig2}(b) the spectrum when both the ordinary lasing and the superradiant lasing occur simultaneously, we can notice that the ordinary lasing frequency is pulled to or pushed away from that of the superradiant lasing. We also notice frequency components corresponding to the sum and difference of those two frequencies, originating from the cross terms such as $J^x s_j^z$ and $J^y s_j^z$ in $\dot{s_j^x}$ and $\dot{s_j^y}$ equations. 

The pulling and pushing of the ordinary lasing frequency by the superradiant lasing component results in a flat region as indicated by a rectangle in the power spectral density plot of Fig.~\ref{fig2}(c). In this region, the lasing frequency is independent of the cavity-atom detuning, exhibiting a zero frequency pulling coefficient. It implies that the lasing frequency would not be affected by the fluctuations in the cavity frequency. Such zero frequency pulling effect is a desired asset in optical clocks. However, the lasing frequency also exhibits a constant offset, not locked to the atomic resonance frequency. Since the offset depends on the initial atomic states in a complicated way, the zero frequency pulling seen in Fig.~\ref{fig2}(c) cannot be used in optical clocks. 

It might be possible, however, to overcome this problem by introducing another superradiant lasing component with the opposite frequency detuning. If two superradiant lasing components at $\pm \Delta\_{pa}$ symmetrically affect the ordinary lasing component in the middle, we might be able to lock the ordinary lasing frequency at the atomic resonance regardless of the cavity-atom detuning.

In order to check the effectiveness of this idea, 
we now prepare the initial superposition state of atom according to the second case discussed in Sec.~\ref{sec-II-B-2} and perform numerical simulations. The results are summarized in Fig.~\ref{fig3}. Only difference among three cases there is the atom-cavity interaction times $\tau$, which are 0.36 $\mu$s, 0.40 $\mu$s and 0.44 $\mu$s in (a), (b) and (c), respectively. The central lasing peak is sandwiched between two superradiant lasing peaks at $\pm\Delta\_{pa}$ around the atomic resonance. Because of the symmetric interaction with the two opposite superradiant lasing peaks, the central lasing peak is lock to the atomic resonance regardless of the cavity-atom detuning, i.e., zero frequency pulling, as best seen in  Fig.~\ref{fig3}(b). The zero frequency pulling occurs over a range of the cavity-atom detuning comparable to the cavity decay rate, which is about 50 MHz in our simulation. Clean central peak is obtained when the empirical condition \hypertarget{Q2}{\new{$\Delta\_{pa}\tau\simeq 5$}} is satisfied with $\tau$ the atom-cavity interaction time.

As to be seen in Fig.~\ref{fig4}, the central peak corresponds to the ordinary lasing whereas the two symmetric side peaks correspond to the superradiant lasing with their linewidth the same as the pump laser linewidth. 
As discussed in Sec.~\ref{secII-C}, the power spectral density of the cavity field is obtained by Fourier transforming the first-order correlation function of $\langle \hat{a}^\dag(t_2)\hat{a}(t_1)\rangle\propto J^+(t_2)J^-(t_1)$ with $t_1$ and $t_2$ chosen after the steady-state is reached. We varied the mean number $N$ of atoms in the cavity and the pump pulse area $\Omega\tau\_p$ independently and calculated the power spectral density. The results are summarized in Fig.~\ref{fig4}. The vertical axis corresponds to the logarithm of the central-peak linewidth in MHz. Over the range of $N$ and $\Omega\tau\_p$ varied, the linewidth changes by 4 orders of magnitude, reaching the minimum of about 50 Hz when $N\simeq 10,000$ and $\Omega\tau\_p\simeq 0.94\pi$. Since the collective emission rate is given by $\Gamma\_c=4g^2/\kappa$, it is natural to compare the linewidth with $\Gamma\_c$.  We note that the minimum linewidth amounts to 1/100 of $\Gamma\_c$ under our simulation conditions. 
In contrast, the two side peaks at $\pm \Delta\_{pa}$ correspond to superradiant lasing and their linewidths are the same as that of the pump field. It is because the superradiant peaks coherently result from initially injected $s_x^0$ and $s_y^0$ in Eq.~(\ref{eq26}) and the linewidth originates from the frequency uncertainty in $\Delta\_{pa}$, i.e., the linewidth of the pump field.

\section{Discussion}

\subsection{Evolution of linewidth reduction}

When the linewidth of the central peak in the power spectral density of the cavity field is plotted as a function of the mean number of atoms in the cavity as in Fig.~\ref{fig4}(b), several distinct features can be noticed. First, the linewidth rapidly drops from an order of MHz, corresponding to the transit time broadening, to about 20 kHz, which is close to the pump-field linewidth in our simulations. This drop is basically the gain narrowing in the conventional laser, the linewidth inversely proportional to the mean number of photons in the cavity. When the linewidth is plotted as a function of the mean photon number in the log-log scale as in Fig.~\ref{fig5}, we find the slope is close to -1. 

Second, the linewidth is further decreased with the mean number of atoms, exhibiting 
a discontinous slope change around the pump linewidth in its decrease. The reduction of the linewidth beyond the pump linewidth is qualitatively different from the gain narrowing in that the mean photon number is clamped or even reduced while the linewidth drops, corresponding to near vertical segments in Fig.~\ref{fig5}. During this drop, the linewidth gets narrowed while the peak height increased and thus the total number of photons remaining about the same. 

The linewidth does not drop forever. Eventually, the linewidth starts to grow, so does the mean photon number, as the number of atoms is increased. This linewidth broadening must be related to power broadening or dressed-state interactions under the \hypertarget{Q3}{\new{strong}} intracavity field, making the atoms no longer a clean two-level system. 

In our simulations, the significant linewidth reduction, down to 100 Hz or less,  takes place in a narrow range of the pump pulse area corresponding to $0.90 < \rho\_{ee}<0.98$, judging from Fig.~\ref{fig5}. 
It is important to have just enough amount of $s_x^0$ and $s_y^0$. Too much or too little $s_x^0$ and $s_y^0$ will make either the ordinary lasing component or the superradiant lasing components too small to induce sufficient interaction between them.
For those pulse areas, the minimum linewidth is obtained with $N\_{min}\sim 8000$ with $\Gamma\_c/2\pi=2g^2/(\pi\kappa)=5$ kHz, so the number $\mathcal{N}\_{coll}$ of collective emission events is given by $\mathcal{N}\_{coll}=N\_{min}\Gamma\_c\tau\sim 100$. One then expects the minimum linewidth to be given by $(\Gamma\_c/2\pi)/\mathcal{N}\_{coll}=1/(2\pi N\_{min}\tau)\sim 50$ Hz, which is exactly what we observe. 

From Fig.~\ref{fig4}(b), one can also note that the linewidth reduction beyond the gain narrowing starts to occur when the number $\mathcal{N}\_{coll}$ of collective emission events becomes much larger than unity. So the linewidth reduction of the central conventional lasing peak beyond the gain narrowing can be thought as being taking place when the conventional lasing evolves to superfluorescence, the collective interaction starting from pure population inversion.

\subsubsection{Collective interaction to induce the minimum linewidth}

Note $1/(2\pi \tau)\sim 390$ kHz corresponds to the transit-time broadening (within a factor of two), so the minimum linewidth can be thought as the transit time broadening divided by the number of atoms. This interpretation is analogous to the linewidth in laser mode locking, where each pulse has a broad linewidth but the each spectral peak has a linewidth reduced by the factor equal to the number of pulses in the pulse train. Likewise, the emission from each atom in our case would exhibit a transit-time broadening but the collective emission of $N\_{min}$ such atoms results in a linewidth reduced from the transit-time broadening by a factor equal to the number $N\_{min}$ of atoms.

In order to reduce the minimum linewidth, one temps to reduce the transit time broadening and increase the number of atoms corresponding to the minimum. But they are not independent from each other, and therefore such adjustment is not possible. We rather pay attention to the empirical fact that 
the minimum linewidth is obtained \hypertarget{Q4}{\new{when $\mathcal{N}\_{coll}\sim 10^2$}}, so the minimum linewidth is roughly given by $\Gamma\_c/100=g^2/(25\kappa)$, suggesting that smaller $g$ and larger $\kappa$ would reduce the minimum linewidth further in the expense of the larger mean number of atoms.

\subsection{Effect of pump carrier detuning and its compensation for an active optical clock}

In practice, the frequency of the pump laser itself may change. If the frequency of the central peak changes in accordance to that of the pump, it is necessary to have a non-frequency indicator with which we can form a feedback loop to correct the pump frequency detuning for optical clock applications. 
For this purpose, we investigate the power spectral density as a function of the pump carrier detuning $\delta$. The pump laser is detuned from the atomic resonance by $\delta$ (carrier detuning) while its amplitude is modulated as $\cos(\Delta\_{pa}t)$, leading to two pump detuning frequencies at $\delta\pm\Delta\_{pa}$. 
The initial conditions for $s_\alpha^0$ ($\alpha=x,y,z$) are derived in Eq.~(\ref{eq32}). The simulation results are summarized in Fig.~\ref{fig6}, which shows in (b) that the three peak spectrum shifts as a whole by the carrier detuning $\delta$. Nonzero $\delta$ also induces imbalance in the side peak heights. The superradiant peak the closer to the atomic resonance becomes the larger whereas the other one the smaller. It is understandable since the gain gets larger toward the atomic resonance.

In applications, the carrier detuning tends to be minimized, so we varied $\delta$ in a small range from -0.03 MHz to +0.03 MHz, within 1.5\% of the modulation frequency $\Delta\_{pa}=2$ MHz. In Fig.~\ref{fig6}(a), the contrast ratio of the peak heights, defined as $(h_>-h_<)/(h_>+h_<)$ with $h_{>(<)}$ the height of the superradiant lasing peak at the higher(lower) detuning, is plotted as a function of $\delta$. We are interested in the contrast ratio for large number of atoms, for which the linewidth of the central peak is greatly reduced. The slope of the contrast ratio is saturated as the number of atoms  is increased, resulting in a slope of about -1/MHz. This height difference can be used as a non-frequency indicator for a feedback loop.
 
\subsubsection{Pump stability with the feedback by peak height difference}
The pump laser stability achievable by a feedback loop can be estimated as follows. Suppose we have a pump carrier detuning of $\delta$. The peak height difference would then be $-2\delta$/(1 MHz) times the mean peak height. To resolve this difference, we need a signal-to-noise ration larger than (1 MHz)/(2$\delta$). Conversely, if the signal-to-noise ratio is equal to (1 MHz)/(2$\delta$), the frequency stability of the pump laser by the feedback mechanism would be just $2\delta$ (full width). If we want the pump stability comparable to the minimum linewidth in Fig.~\ref{fig5}, the signal-to-noise ratio should be (1 MHz)/(50 Hz)$=2\times10^4$. The output photon flux is estimated as $n \kappa\sim 3n\times 10^{8}$ photons/sec according to Fig.~\ref{fig5} with \hypertarget{Q5}{\new{$n$ the number of photons associated with the average of the side peak heights}}.  
For an averaging time of $T$, we get $3n(T/{\rm sec})\times10^{8}$ photons, and thus the signal-to-noise ratio is about $\sqrt{3n(T/{\rm sec})}10^4$, and thus an enough signal-to-noise ratio is obtained if $n>1.3/(T/{\rm sec})$. The condition is easily satisfied in our simulations for $T\sim 1$ sec, the pump laser stability can be made comparable to the minimum linewidth of the central peak. As a whole, the pump laser and our laser may then form an active optical clock.

\section{Conclusion}
In this paper, we considered a beam of two-level atoms traversing a low-Q cavity. The atoms were pre-pumped to a superposition state of the ground and excited states with the excited-state amplitude much larger than that of the ground state before they enter the cavity. We assumed a traveling-wave interaction between the atoms and the cavity and the mode matching between the pump and the cavity mode.
We numerically solved the quantum Langevin equations when the Rabi frequency of the pump laser at atomic resonance was modulated at $\Delta\_{pa}$ for various cavity-atom detuning. We found that the conventional lasing occurred at the atomic resonance and simultaneously superradiant lasing took place at $\pm\Delta\_{pa}$ around the atomic resonance regardless of the cavity-atom detuning for both, resulting in a zero frequency pulling coefficient for the conventional lasing coming from the initial population inversion. In addition, the linewidth of the central lasing peak was reduced beyond the gain narrowing while the mean photon number is stationary as the mean number of atoms was increased. The transition occurred when the number of collective emission events becomes much larger than unity, signaling the conventional lasing transforming to superfluoresce. The minimum linewidth, as small as a millionth of the atomic natural linewidth or the cavity linewdith, was given by the transit time broadening for individual atoms divided by the number $N$ of atoms associated with the minimum, analogous to the spectral linewidth of $N$-mode laser mode locking. When the pump carrier frequency was slightly detuned by $\delta$ from the atomic resonance, the three lasing peaks as a whole was shifted by $\delta$ with asymmetric heights for the side peaks. According to our signal-to-noise ratio analysis, it would be possible to lock the pump laser to the atomic resonance with its uncertainty comparable to the linewidth of the central peak by employing a simple feedback loop based on the peak height difference. Our results of the zero frequency pulling as well as the narrow linewidth are attractive features for clock applications and thus may lead to a new type of active optical clocks for future frequency standards when applied to proper atomic systems.

\acknowledgements
We thank Myoungseon Hur for helpful discussions on eliminating the pump laser detuning. This work was supported by National Research Foundation of Korea (Grant No.~2020R1A2C3009299).

\bibliographystyle{naturemag}
\bibliography{ref-rev}


\end{document}